\documentclass[twocolumn,trackchanges]{aastex63}

\usepackage{graphicx}
\graphicspath{{./figs/}}
\usepackage{amsmath}
\usepackage[encapsulated]{CJK}
\usepackage{ucs}

\newcommand{\eazy}{\texttt{EAzY}}
\newcommand{\prospector}{\texttt{Prospector}}
\newcommand{\pb}{\texttt{Prospector}-$\beta$}

\newcommand{\msun}{{\rm M}_{\odot}}
\newcommand{\mc}{$M\rm{_c}$}

\shorttitle{Prospector-$\beta$: Survey Mode}
\shortauthors{Wang et al.}

\begin{document}

\title{Inferring More from Less: Prospector as a Photometric Redshift Engine in the Era of JWST}

\correspondingauthor{Bingjie Wang}
\email{bwang@psu.edu}

\author[0000-0001-9269-5046]{Bingjie Wang (\begin{CJK*}{UTF8}{gbsn}王冰洁\ignorespacesafterend\end{CJK*})}
\affiliation{Department of Astronomy \& Astrophysics, The Pennsylvania State University, University Park, PA 16802, USA}
\affiliation{Institute for Computational \& Data Sciences, The Pennsylvania State University, University Park, PA 16802, USA}
\affiliation{Institute for Gravitation and the Cosmos, The Pennsylvania State University, University Park, PA 16802, USA}

\author[0000-0001-6755-1315]{Joel Leja}
\affiliation{Department of Astronomy \& Astrophysics, The Pennsylvania State University, University Park, PA 16802, USA}
\affiliation{Institute for Computational \& Data Sciences, The Pennsylvania State University, University Park, PA 16802, USA}
\affiliation{Institute for Gravitation and the Cosmos, The Pennsylvania State University, University Park, PA 16802, USA}

\author[0000-0001-5063-8254]{Rachel Bezanson}
\affiliation{Department of Physics \& Astronomy and PITT PACC, University of Pittsburgh, Pittsburgh, PA 15260, USA}

\author[0000-0002-9280-7594]{Benjamin D. Johnson}
\affiliation{Center for Astrophysics $\vert$ Harvard \& Smithsonian, Cambridge, MA 02138, USA}

\author[0000-0002-3475-7648]{Gourav Khullar}
\affiliation{Department of Physics \& Astronomy and PITT PACC, University of Pittsburgh, Pittsburgh, PA 15260, USA}

\author[0000-0002-2057-5376]{Ivo Labb\'{e}}
\affiliation{Centre for Astrophysics and Supercomputing, Swinburne University of Technology, Melbourne, VIC 3122, Australia}

\author[0000-0002-0108-4176]{Sedona H. Price}
\affiliation{Department of Physics \& Astronomy and PITT PACC, University of Pittsburgh, Pittsburgh, PA 15260, USA}

\author[0000-0003-1614-196X]{John R. Weaver}
\affiliation{Department of Astronomy, University of Massachusetts, Amherst, MA 01003, USA}

\author[0000-0001-7160-3632]{Katherine E. Whitaker}
\affiliation{Department of Astronomy, University of Massachusetts, Amherst, MA 01003, USA}
\affiliation{Cosmic Dawn Center (DAWN), Niels Bohr Institute, University of Copenhagen, Jagtvej 128, K{\o}benhavn N, DK-2200, Denmark}

\begin{abstract}

The advent of the James Webb Space Telescope (JWST) signals a new era in exploring galaxies in the high-$z$ universe. Current and upcoming JWST imaging will potentially detect galaxies out to $z \sim 20$, creating a new urgency in the quest to infer accurate photometric redshifts (photo-$z$) for individual galaxies from their spectral energy distributions, as well as masses, ages and star formation rates. Here we illustrate the utility of informed priors encoding previous observations of galaxies across cosmic time in achieving these goals. We construct three joint priors encoding empirical constraints of redshifts, masses, and star formation histories in the galaxy population within the \prospector\ Bayesian inference framework. In contrast with uniform priors, our model breaks an age-mass-redshift degeneracy, and thus reduces the mean bias error in masses from 0.3 to 0.1 dex, and in ages from 0.6 to 0.2 dex in tests done on mock JWST observations. Notably, our model recovers redshifts at least as accurately as the state-of-the-art photo-$z$ code \eazy\ in deep JWST fields, but with two advantages: tailoring a model based on a particular survey renders mostly unnecessary given well-motivated priors; obtaining joint posteriors describing stellar, active galactic nuclei, gas, and dust contributions becomes possible. We can now confidently use the joint distribution to propagate full non-Gaussian redshift uncertainties into inferred properties of the galaxy population. This model, ``\prospector-$\beta$'', is intended for fitting galaxy photometry where the redshift is unknown, and will be instrumental in ensuring the maximum science return from forthcoming photometric surveys with JWST. The code is made publicly available online as a part of \prospector\ \footnote{The version used in this work corresponds to the state of the Git repository at commit \href{https://github.com/bd-j/prospector/commit/820ad72363a1f9c22cf03610bfe6e361213385cd}{\texttt{820ad72}}.}.

\end{abstract}


\section{Introduction}

Within months of the first data releases, JWST has already begun to revolutionize our view of galaxy formation (e.g., \citealt{2022arXiv221105792F,Treu2022,2023AJ....165...13W}). Fitting spectral energy distributions (SEDs) plays a key role in grounding new observations to theories of galaxy evolution.
State-of-the-art Bayesian codes including \texttt{BAGPIPES} \citep{Carnall2018}, \texttt{BEAGLE} \citep{Chevallard2016}, and \prospector\ \citep{Johnson2021}, have been applied extensively in this context.
One of the most advanced models in this family is \prospector-$\alpha$ \citep{Leja2017}. It incorporates galaxy properties including nonparametric star formation histories (SFHs; \citealt{Leja2019}), self-consistent nebular emission \citep{Byler2017}, and a sophisticated dust model (e.g., \citealt{Lower2022}), and critically, fits all parameters at once with Markov chain Monte Carlo or nested sampling to produce joint high-dimensional constraints.
\prospector-$\alpha$ has been designed specifically to extract the most information from high signal-to-noise photometry or spectra with good wavelength coverage. Therefore, it has been traditionally applied to well-sampled SEDs using redshifts measured externally, e.g., from spectra or photo-$z$ codes.

With regard to photo-$z$ codes, considerable efforts have been devoted to various developments (see \citealt{Salvato2019,Newman2022} for recent reviews, and also \citealt{Alsing2022,Leistedt2022} for general discussions on photo-$z$ frameworks). The most common algorithm is inferring redshift by comparing observations to SED templates, as is used in \texttt{LePhare} \citep{Arnouts1999,Ilbert2006}, \texttt{BPZ} \citep{Benitez2000}, \texttt{ZEBRA} \citep{Feldmann2006}, \eazy\ \citep{Brammer2008}, and others. This method suffers from the fact that template colors are frequently degenerate with redshift. To mitigate this problem, some codes add on a magnitude-dependent redshift prior (e.g., \texttt{BPZ}, \eazy), the probability distribution of which is determined by fitting redshift distributions of existing catalogs in magnitude bins. However, few codes have adopted a full Bayesian approach, in which priors can be assigned to every parameter. Conversely, codes that are capable of mapping the multi-dimensional posterior probability distributions for redshift and galaxy properties simultaneously via advanced sampling techniques, as mentioned earlier, are not optimized for fitting SEDs when the redshift is unknown.

The era of JWST creates an urgent new opportunity to measure the highly uncertain early phases of galaxy formation, and, for the first time, necessitates allowing redshift solutions over the range of $0 < z \lesssim 20$.
Thus there is an increasing need for accurately inferring  joint, correlated constraints between redshift and important galaxy parameters like mass, star formation rate (SFR), and dust properties where exquisite wavelength coverage and/or external spectra are lacking.

In this work, we establish a framework in which empirical constraints of galaxy evolution are encoded directly into the inference process as informative priors, so that redshift and galaxy properties can be constrained simultaneously and accurately.
It is a known challenge that the photo-$z$ accuracy depends strongly on the assumptions that go into the code (e.g., \citealt{2023ApJ...942...36K}).
Here we propose that the most accurate assumptions are our previous observations.
A common argument for assuming uniform priors when fitting for high-$z$ objects is that the epoch in question is largely unexplored and hence uniform priors are the least sinful (e.g., \citealt{Finkelstein2022}). Here we argue for the opposite: priors represent our belief about which solutions we think are more probable, and it is thus always preferable to use well-motivated priors if available.

We present a new model, \pb, optimized to recover photo-$z$ in deep JWST fields, while taking full advantage of the capability of \prospector\ to produce a high-dimensional SED-model and obtain joint constraints on all inferred physical parameters.
This means that the full probability distribution can be used to propagate full non-Gaussian redshift uncertainties into inferred properties of the galaxy population. Doing so will significantly enhance our confidence in the inferred properties, and will thus maximize the information returned from JWST.

We devise three new priors: a mass function prior, a galaxy number density prior, and a dynamic non-parametric SFH prior which reflects the consistent observational finding that massive galaxies form much earlier than low-mass galaxies \citep{Cowie1996,Thomas2005}. Our SFH prior also respects the observed cosmic star formation rate density by encouraging rising histories early in the universe, and falling histories late in the universe.

Moreover, we identify and characterize an age-mass-redshift degeneracy that contaminates the results of standard uniform priors. We show that our model is able to break this degeneracy, while recovering redshifts at least as accurately as \eazy\ in JWST surveys.

The letter is structured as follows. Section~\ref{sec:data} states the simulated and the observed photometry on which we assess our model performance. Section~\ref{sec:priors} details the construction of the three observationally-informed priors. Section~\ref{sec:fit} describes the procedure of SED fitting. Section~\ref{sec:res} reports the results. Section~\ref{sec:concl} summarizes our findings and concludes with implications for future works. When applicable, we adopt the best-fit cosmological parameters from the nine-year results from the Wilkinson Microwave Anisotropy Probe mission: $H_{0}=69.32$ ${\rm km \,s^{-1} \,Mpc^{-1}}$, $\Omega_{M}=0.2865$, and $\Omega_{\Lambda}=0.7135$ \citep{Bennett2013}.

\section{Data\label{sec:data}}

While the framework devised in this letter is applicable to any data, we scope the current work to JWST surveys in light of the immediate need to explore the parameter space that has just been opened by JWST. With this in mind, we construct our simulation imitating a JWST survey to quantify the effect of the proposed priors.

The mock catalog is based on JAGUAR, which describes the evolution of galaxies across $0 \leq z \leq 15$ and $7 \leq {\rm log} M_\star/\msun \leq 13$ \citep{Williams2018}. It uses observed stellar mass and UV luminosity functions at $0 < z < 10$ \citep{Bouwens2015,Oesch2018} to model the evolution of stellar mass, UV absolute magnitude, UV continuum slope, and galaxy number counts. The relevant parameters then become the inputs to \texttt{BEAGLE} \citep{Chevallard2016} to generate SEDs in JAGUAR.

We note that \texttt{BEAGLE} and \prospector\ incorporate different stellar population synthesis: the former uses \cite{Bruzual2003}, while the latter calls FSPS \citep{Conroy2010,Choi2016,Dotter2016}. The underlying nebular emission, stellar templates, and isochrones all differ. In order to isolate the effect of the prior and avoid concerns about template differences, we re-construct the mock catalog using identical inputs for FSPS.
The stellar populations are modeled with a Chabrier initial mass function \citep{Chabrier2003}, and a two-component dust model \citep{Charlot2000}. The SFH is parameterized by a delayed-$\tau$ model.

We add noise typical for a deep field JWST + HST survey, then impose a $z\geq 0.5$ cut and a signal-to-noise $\geq 5$ cut based on the flux in the F444W band, which result in about 56,\,000 mock galaxies. Further details on the mock catalog can be found in the Appendix.

In addition to the simulations, we test our model on JWST data as well. The observations are recently acquired in a Director's Discretionary (DD) program (JWST-DD-2756, PI: Chen). We cross-match the sources to collect spectroscopic redshifts from the NASA/IPAC Extragalactic Database, \citet{Treu2015}, and \citet{Richard2021}. Data reduction follows \citet{Weaver2023}.
The observed fluxes in the F444W band span roughly from 16.5 to 30.5 AB mag, with a 5$\sigma$ depth of 25.5 AB mag. Star-forming galaxies constitute the majority of this sample.

\section{Galaxy Evolution Priors\label{sec:priors}}

\begin{figure*}
\gridline{
    \fig{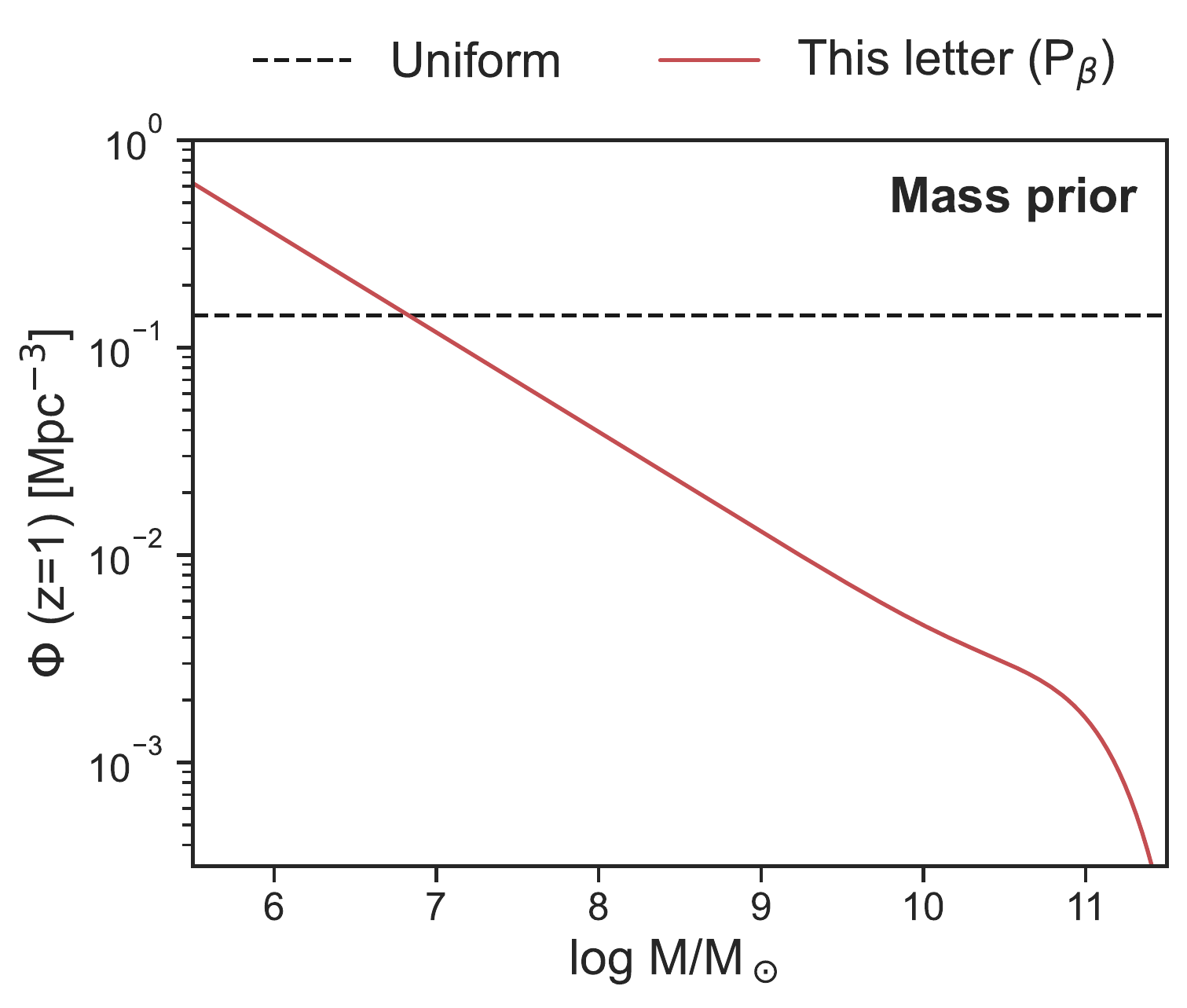}{0.33\textwidth}{(a)}
    \fig{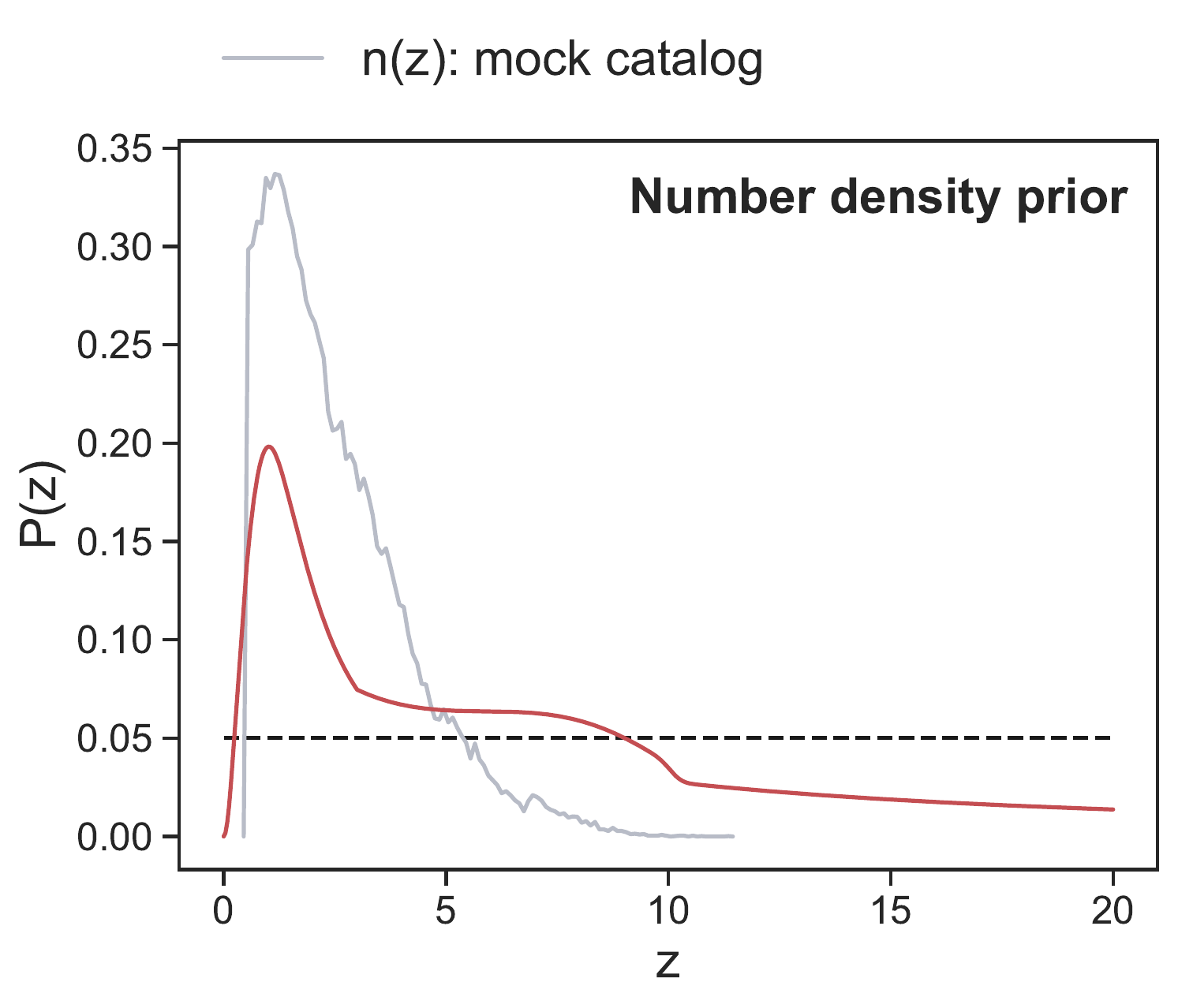}{0.33\textwidth}{(b)}
    \fig{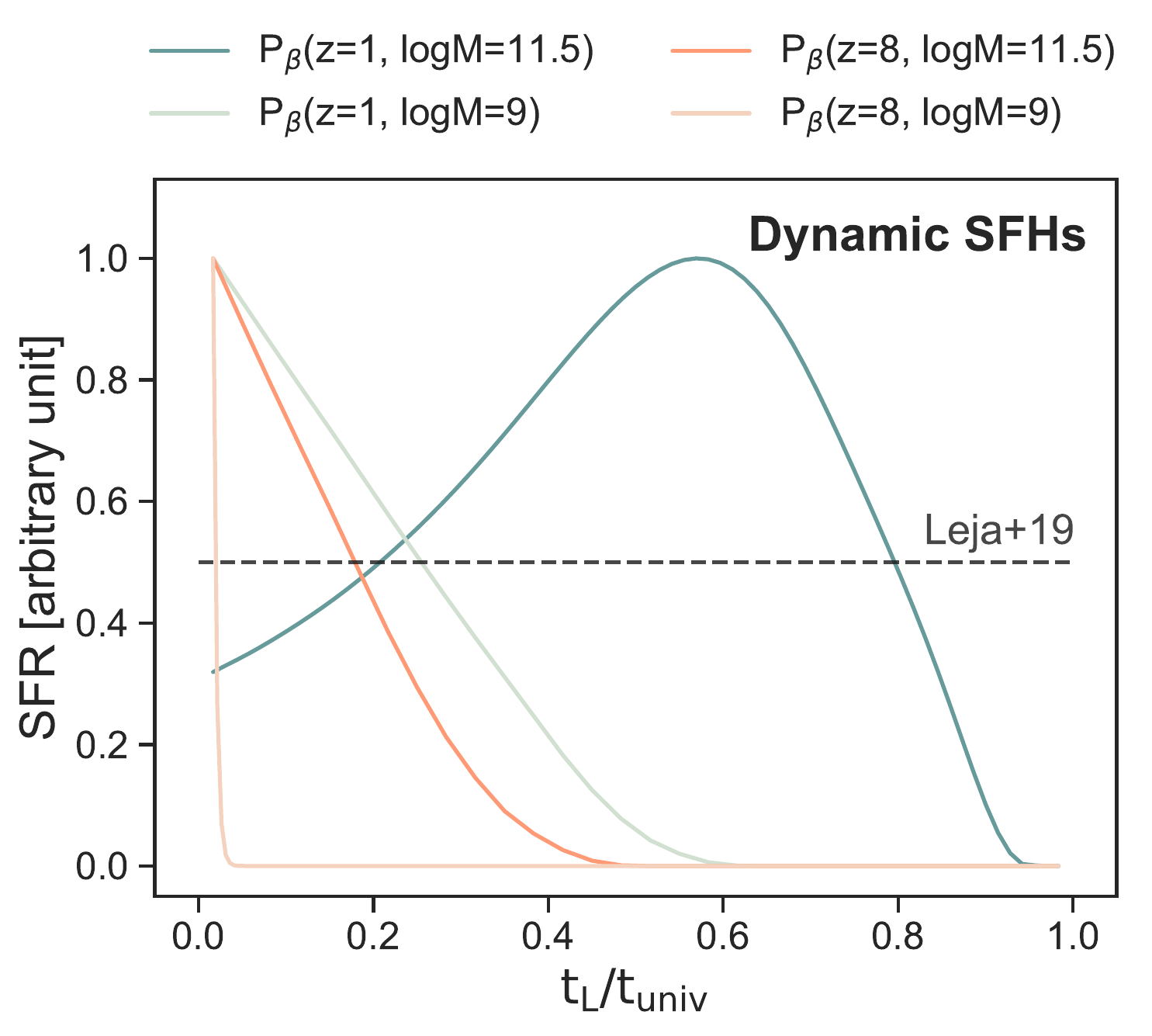}{0.33\textwidth}{(c)}
    }
\caption{The three panels illustrate the difference between the proposed priors on mass, number density, and SFH of this letter, and uniform priors (black dashed lines). (a) Mass function at $z=1$ \citep{Leja2020} is shown in red. (b) Observable galaxy number density (i.e., $M_c$ cut applied), as a function of redshift is plotted in red. Also shown in gray is the number density directly calculated from the mock catalog, which differs from our $P(z)$ due to the different assumptions built in; in particular, mass functions at $z \gtrsim 5$. Nevertheless, we are able to improve the parameter inference by assuming a reasonable approximation of the mass functions used in the mock catalog. (c) Four SFHs from our prior are shown as colored lines. The x-axis is lookback time, $t_{\rm L}$, normalized by the age of the universe, $t_{\rm univ}$, at the respective $z$, and the y-axis is also normalized so that all SFRs can be shown on the same scale.\label{fig:prior}}
\end{figure*}

\begin{figure*}
\gridline{
    \fig{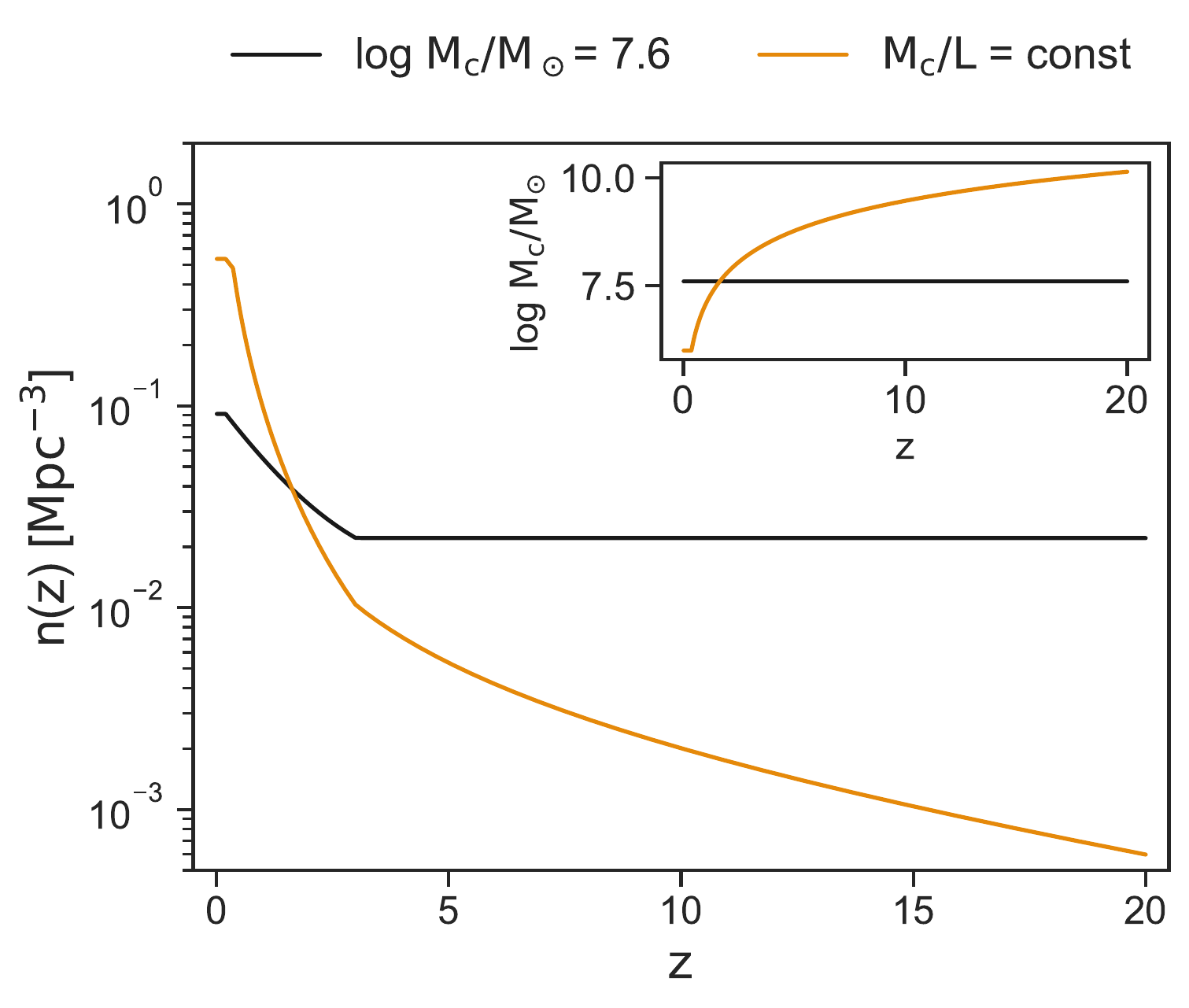}{0.33\textwidth}{(a)}
    \fig{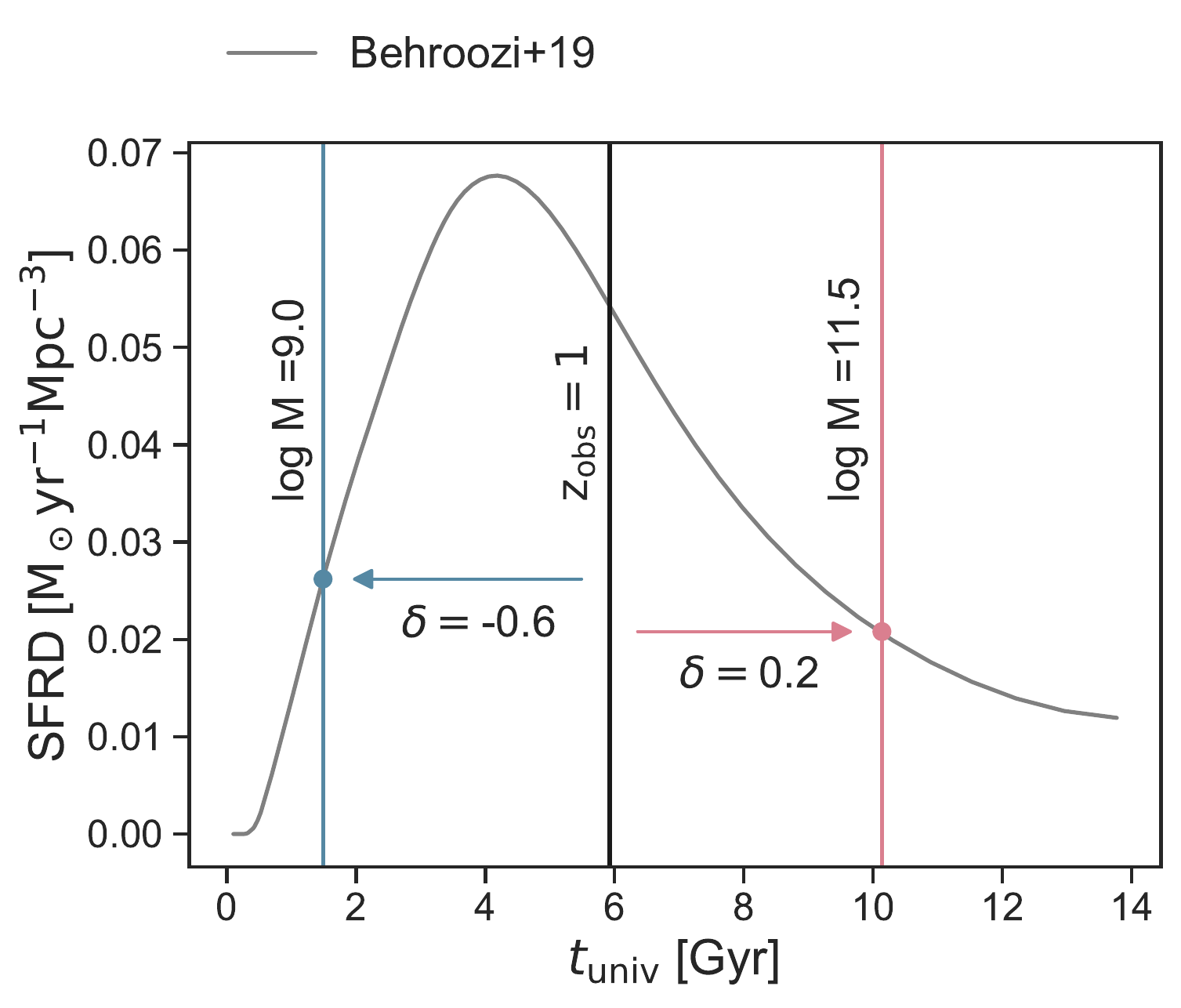}{0.33\textwidth}{(b)}
    \fig{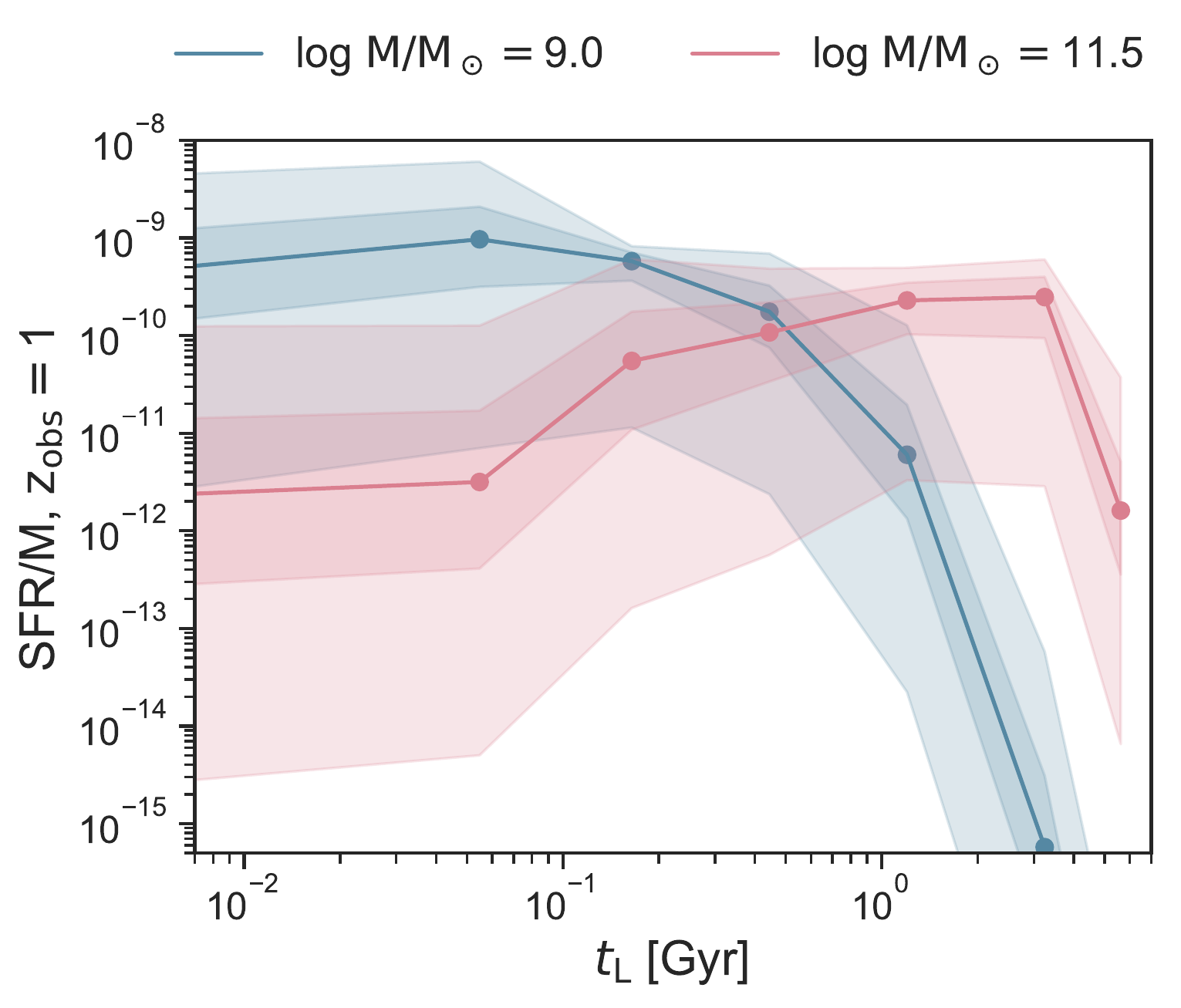}{0.33\textwidth}{(c)}
}
\caption{Here we show the technical details of the implementation of our priors from observations of galaxy evolution. (a) Number densities of galaxies, $n$, assuming different evolutions of the mass complete limit, \mc, are plotted as functions of redshift. Shown in black is $n$ based on a constant \mc, while shown in dark orange is based on a constant mass-to-light ratio. The inserted panel shows \mc\ as a function of $z$. (b) A schematic diagram demonstrating our $\delta$-formalism, which introduces mass dependence to the SFH prior. The cosmic SFRD \citep{Behroozi2019}, which we take to be the expectation value, is plotted as a function of the age of the universe in gray. At an observed redshift, $z_{\rm obs}$, the expected SFR values of a galaxy are shifted back/forward in time depending on its mass. The constructed SFH prior favors rising SFHs in the early universe, and falling SFHs in the late universe. (c) Specific SFRs of two galaxies of masses $10^8\msun$ and $10^{10.5}\msun$, observed at $z=1$, are plotted as functions of lookback time in blue and red, respectively. The x-axis is the time between the onset of star formation (birth of the galaxy) and $z=1$. The shades indicate the 16--84th and the 2.5--97.5th percentiles of the prior distribution. \label{fig:sfh}}
\end{figure*}

In this section, we detail the construction of three observationally-informed priors on galaxy stellar mass, the observable number density, and galaxy star formation history SFH$(M,z)$. We contrast each with the respective uniform prior in Figure~\ref{fig:prior}.

\subsection{Stellar Mass Function\label{sec:pmz}}

Galaxy stellar mass can be strongly constrained from photometry when the distance is known \citep{Bell2001,Pforr2012,Conroy2013,Mobasher2015,Li2022}; however when the distance is unknown, inferring mass becomes challenging (mass refers to stellar mass hereafter).
An informative prior can be helpful in this case. In nature, low-mass galaxies are far more numerous than high-mass galaxies, so a mass function prior allows us to avoid spurious high-mass, high-$z$ solutions. For a given $z$, we draw mass from ${\rm P(M} | z)$, which is just a mass function, $\Phi$, at that redshift normalized such that
\begin{equation}
	\int_{M_{\rm min}}^{M_{\rm max}} \Phi (z) dM = 1,
	\label{eq:pmz}
\end{equation}
where $M_{\rm min}$ and $M_{\rm max}$ are the minimum and maximum masses. We set $M_{\rm min}=10^{6} \msun$ and $M_{\rm max}=10^{12} \msun$ in this work.

In this letter, we take the mass functions from \cite{Leja2020}, which are continuous in redshift and are derived using the same stellar population model, \prospector-$\alpha$.
We note that they are only defined between $0.2 \leq z \leq 3.0$. For $z < 0.2$ and $z > 3$, we adopt a nearest-neighbor solution, i.e., the $z=0.2$ and $z=3$ mass functions.
The nearest-neighbor choice may be favored in the absence of reliable high-resolution rest-frame optical selected mass functions at $z > 3$. Importantly, it allots a conservatively high probability for yet-to-be-discovered populations of high-mass, high-$z$ galaxies, hints of which have already been observed in \citet{Labbe2022}.

Conversely, if the goal is to decrease as much as possible the contamination of interlopers posing as extremely luminous galaxies at $z \gtrsim 10$ \citep{Furlanetto2022}, one may consider transitioning over to theoretical mass functions. To this end, we additionally provide an option in the public release of our code: the mass functions can be switched to those from \cite{Tacchella2018} between $4 < z < 12$, with the $3 < z < 4$ transition from \cite{Leja2020} managed with a smoothly-varying average in number density space. These can easily be replaced with observed mass functions at high $z$ once they become available. While at this stage the high-$z$ mass functions remain uncertainty, we note that even our current estimate is inaccurate, including a mass function prior is a better choice than a flat prior. As can be seen from Figure~\ref{fig:prior}(a), the number density at $10^9 \msun$ at $z=1$ already deviates from the uniform prior by over an order of magnitude.

\subsection{Galaxy Number Density\label{sec:pz}}

The probability of finding a galaxy at redshift $z$ can be estimated from
\begin{equation}
	P(z) = N(z) \, V_{\rm co}(z) = \int_{M_c}^{\infty} \Phi (M, z) \, dM \, V_{\rm co}(z),
	\label{eq:pz}
\end{equation}
where $N(z)$ is the expected total number density of galaxies, $V_{\rm co}(z)$ is the differential co-moving volume, and $M_c$ is the mass completion limit. We normalize $P(z)$ by letting $\int P(z) = 1$. We show $P(z)$ adopted for this work in Figure~\ref{fig:prior}(b).

As shown in Figure~\ref{fig:sfh}(a), the number density, and hence $P(z)$, is sensitive to the mass-completeness limit of the data. This means that a careful analysis must be carried out to estimate $M_{c}(z)$.
In practice one would likely need to obtain $M_c$ from using SED-modeling heuristics based on the flux-completeness limits in a given catalog. Many subtleties exist in such translation due to the complexities of stellar populations (e.g., \citealt{Pozzetti2010,Leja2020}), the discussion of which is beyond the scope of this work.
For simplicity, we extract $M_c$ directly from our mock catalog as the mass above which 90\% of the galaxies in a given redshift bin is included.

\subsection{Dynamic Star-formation History\label{sec:sfh}}

An advantage of \prospector-$\alpha$ is the ability to adapt flexible SFHs. The SFH is set by the mass formed in 7 logarithmically-spaced time bins, assuming a continuity prior which ensures smooth transitions between bins \citep{Leja2019}. While the flat SFH in \prospector-$\alpha$ is a reasonable null expectation, it is inconsistent with expectations from real galaxies: they have rising SFHs in the early universe and falling SFHs at late times, and also show systematic trends with mass (e.g., \citealt{Leitner2012,Pacifici2016}). As pointed out in \citet{Tacchella2022} and \citet{Whitler2022}, the inferred early formation history of high-$z$ galaxies heavily depends on the SFH prior.

We therefore modify the flat SFH assumption by matching the expectation value in each bin in lookback time to the cosmic star formation rate densities (SFRD) in \cite{Behroozi2019}, while still keeping the distribution about the mean identical to the Student's-t distribution in \prospector-$\alpha$.
Taking a step further, we introduce a mass dependence to the SFH prior, since it is well-established that galaxy SFHs depend both on mass and cosmic time \citep{Behroozi2019}. A difficulty, though, is that SFH constraints require highly informative spectroscopy (e.g., \citealt{Carnall2019,Leja2019,Park2022,Tacchella2022a}). In the spirit of keeping the priors minimalistic, we opt to model the mass dependence by shifting the start of the age bins as
\begin{equation}
	\rm log({\it t}_{start}/Gyr) = \rm log({\it t}_{univ}({\it z})/Gyr) + \delta_m,
\end{equation}
where
\begin{equation}
\delta_{\rm m} =
    \begin{cases}
      -0.6, & {\text{for}} \; {\rm log {\it M}/M_\odot} < 9.\\
      1/3 \, \rm log M -3.6, & {\text{for}} \; 9 \leq {\rm log {\it M}/M_\odot} \leq 12.\\
      0.4, & {\text{for}} \; {\rm log {\it M}/M_\odot} > 12.
    \end{cases}
\end{equation}
$\rm {\it{t}}_{univ}({\it z})$ is the age of the universe at an observed $z$. This does not, however, mean that the onset of star formation can be earlier than the Big Bang.
The equation refers to the age of the universe at which the shape of the SFRD is taken as a prior for the expectation values of the time bins. In practice, we shift the start of the age bins along the SFRD curve, and then re-scale the age bins to be within the age of the universe at a given $z$.
This SFH prior effectively encodes an expectation that high-mass galaxies form earlier, and low-mass galaxies form later, than naive expectations from the cosmic SFRD. In other words, the physical motivation is the downsizing scenario in galaxy formation \citep{Cowie1996,Thomas2005}. We show four SFHs from our prior in Figure~\ref{fig:prior}(c) as examples, and technical details of the $\delta$-formalism in Figures~\ref{fig:sfh}(b)-(c).

This prior does not, however, account for the quiescent/star-forming bimodality. We effectively approximate a double-peaked distribution of SFRs at a given mass and redshift seen in, e.g., \citet{Leja2022}, as a wide single-peaked distribution. The quiescent fractions calculated from our prior match roughly with the observed trend at $z < 3$ \citep{Leja2022}, which justifies this simplification.

Having presented the new priors, we note that we include the stellar mass–stellar metallicity relationship measured from the SDSS as a prior as well \citep{Gallazzi2005}. Following \citet{Leja2019}, we take the conservative approach of widening the confidence intervals from this relationship by a factor of 2 to account for potential unknown systematics or redshift evolution.

\section{Fitting Mock Galaxies\label{sec:fit}}

\begin{deluxetable*}{p{0.1\textwidth} p{0.4\textwidth} p{0.4\textwidth}}
\tablecaption{Parameters and Priors for Fitting SEDs within \prospector \label{tab:theta}}
\tablehead{
\colhead{Parameter} & \colhead{Description} & \colhead{Prior}
}
\startdata
$\log (\mathrm{M}/\mathrm{M}_{\odot})$ & total stellar mass formed & see Section~\ref{sec:pmz}\\
$z$ & redshift & see Section~\ref{sec:pz} \\
SFH & ratio of SFRs in adjacent log-spaced time bins & see Section~\ref{sec:sfh}\\
$\log (\mathrm{Z}^{\star}/\mathrm{Z}_{\odot})$ & stellar metallicity & Gaussian approximating the $\mathrm{M}$--Z$^{\star}$ relationship of \cite{Gallazzi2005} \\
$n$ & power law index for a \citet{Calzetti2000} attenuation curve & uniform: $\mathrm{min}=-1.0$, $\mathrm{max}=0.4$ \\
$\hat{\tau}_{\rm dust, 2}$ & optical depth of diffuse dust \citep{Charlot2000} & truncated normal: $\mathrm{min}=0$, $\mathrm{max}=4$, $\mu=0.3$, $\sigma=1$\\
$\hat{\tau}_{\rm dust, 1} / \hat{\tau}_{\rm dust, 2}$ & ratio between the optical depths of birth cloud dust and diffuse dust \citep{Charlot2000} & truncated normal: $\mathrm{min}=0$, $\mathrm{max}=2$, $\mu=1$, $\sigma=0.3$ \\
$\log f_{\mathrm{AGN}}$ & ratio between the object's AGN luminosity and its bolometric luminosity & uniform: $\mathrm{min}=-5$, $\mathrm{max}=\log 3$ \\
$\log \tau_{\mathrm{AGN}}$ & optical depth of AGN torus dust & uniform: $\mathrm{min}=\log 5$, $\mathrm{max}=\log 150$ \\
$\log(\mathrm{Z}_{\mathrm{gas}}/\mathrm{Z}_{\odot})$ & gas-phase metallicity & uniform: $\mathrm{min}=-2.0$, $\mathrm{max}=0.5$ \\
$\mathrm{q_{PAH}}$ & fraction of grain mass in PAHs \citep{Draine2007} & truncated normal: $\mathrm{min}=0$, $\mathrm{max}=7$, $\mu=2$, $\sigma=2$\\
\enddata
\end{deluxetable*}

We fit SEDs with the \prospector-$\alpha$ model, as well as \pb\ of this letter, which essentially replaces the former's uniform priors with informative ones. We emphasize that the physical model shared between \prospector-$\alpha$ and \pb\ is more sophisticated than the one used for generating the mock galaxies. It consists of 16 free parameters describing the contribution of stars, AGNs, gas, and dust (Table~\ref{tab:theta}), although the dust emission and the infrared AGN parameters are only relevant at $z \lesssim 1$.
The posteriors are sampled with the dynamic nested sampling code \texttt{dynesty} \citep{Speagle2020}.
We fit 5,000 randomly drawn mock galaxies. This sample size is limited by the time required of FSPS/nested sampling, which takes $>10$ hours per fit. This problem will be circumvented via machine-learning-accelerated techniques in the near term \citep{Alsing2020,2211.03747}.

For comparison, we determine redshift using \eazy, which fits a non-negative linear combination of a set of templates \citep{Brammer2008}. We use the vanilla version with the default set of \verb|tweak_fsps_QSF_12_v3|.

\section{Results\label{sec:res}}

\begin{figure*}
\gridline{
    \fig{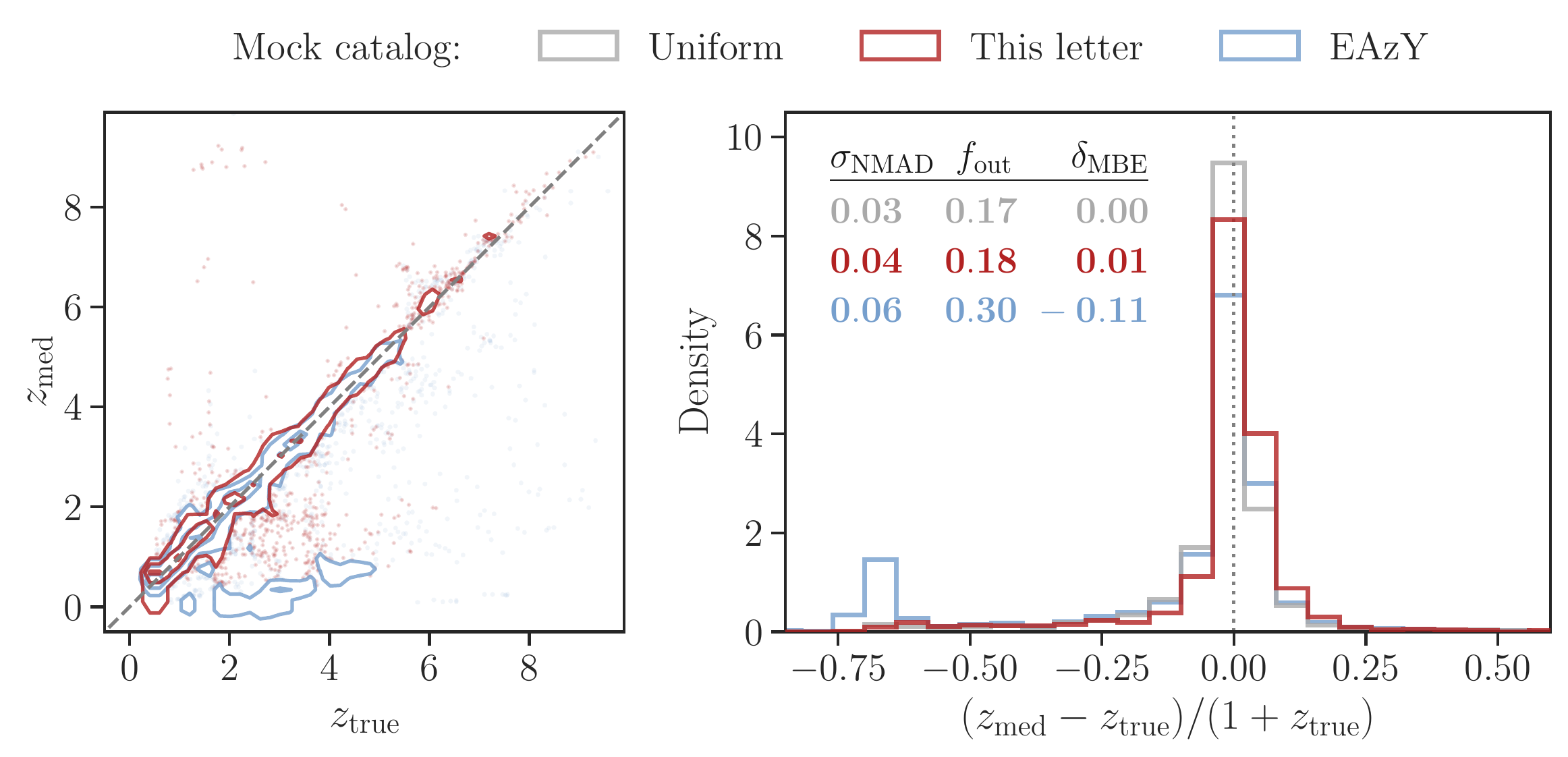}{0.95\textwidth}{}
}
\gridline{
    \fig{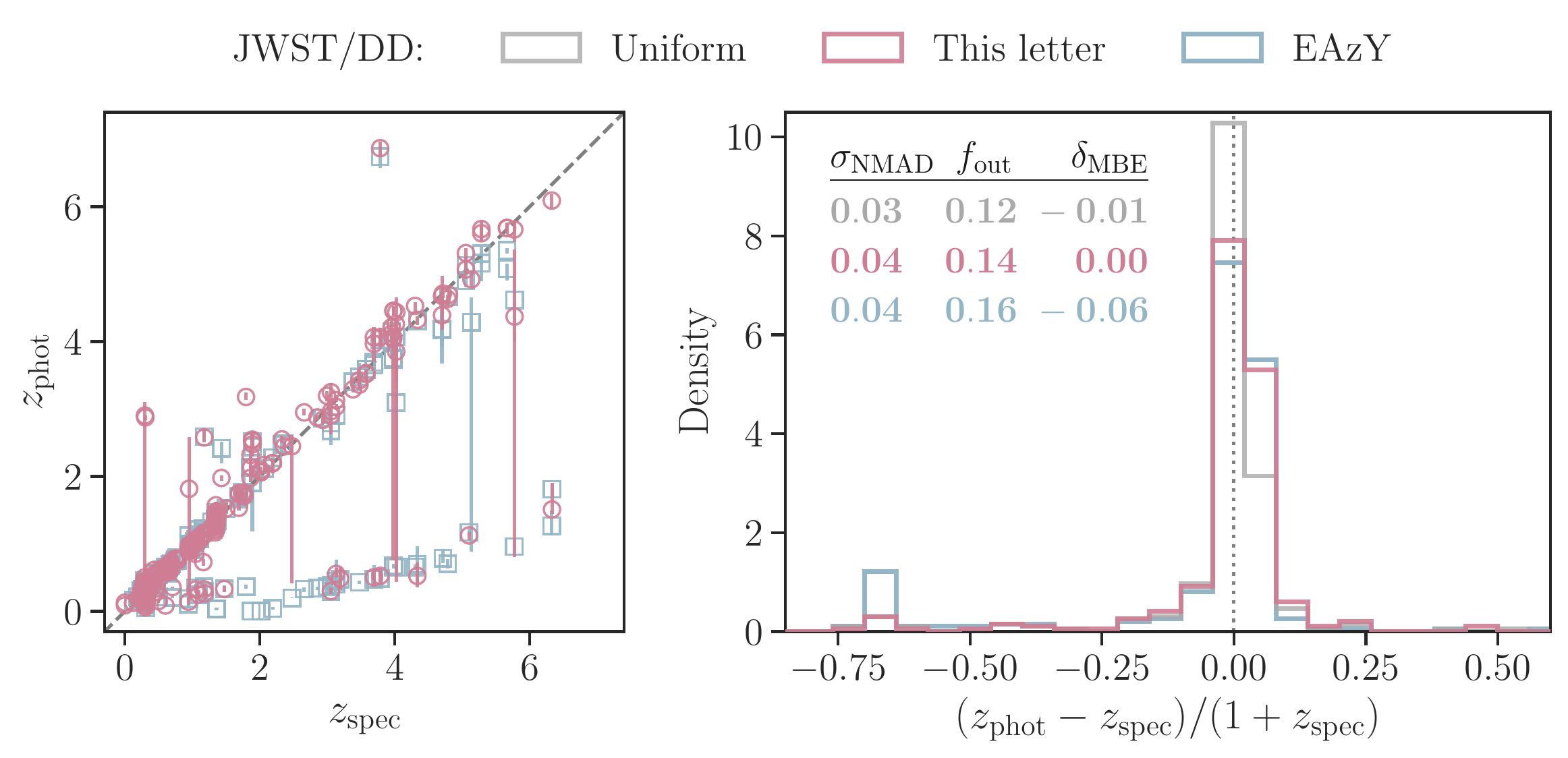}{0.95\textwidth}{}
}
\caption{We compare the photo-$z$ accuracy between \pb\ and \eazy\ using both simulated and observed data.
 (upper panel) On the left, we show true $z$ vs. posterior median $z_{\rm med}$ from a mock catalog. Red contours show results from \pb, while blue contours show these from \eazy. On the right, we show residuals as histograms. We also include results based on uniform priors, plotted as gray histograms. NMAD, outlier fraction, and MBE are reported in the same colors as the corresponding histogram. (lower panel) Same as the upper panel, but for redshifts inferred from a spec-$z$ sample with newly acquired data from a JWST DD program. In both cases, we find that \pb\ performs at least as accurately as \eazy. \label{fig:hist_zred}}
\end{figure*}

\begin{figure*}
\gridline{
    \fig{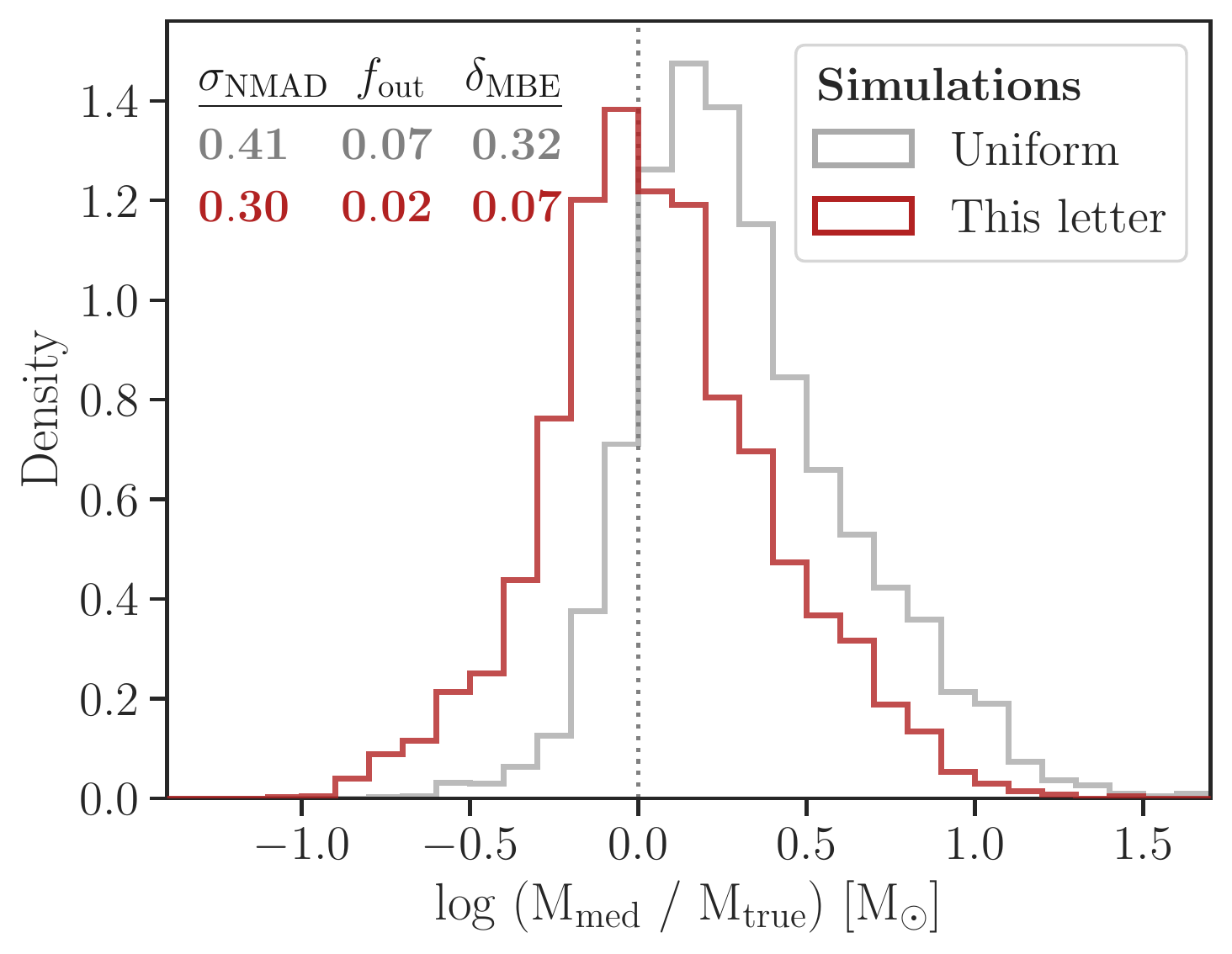}{0.325\textwidth}{}
    \fig{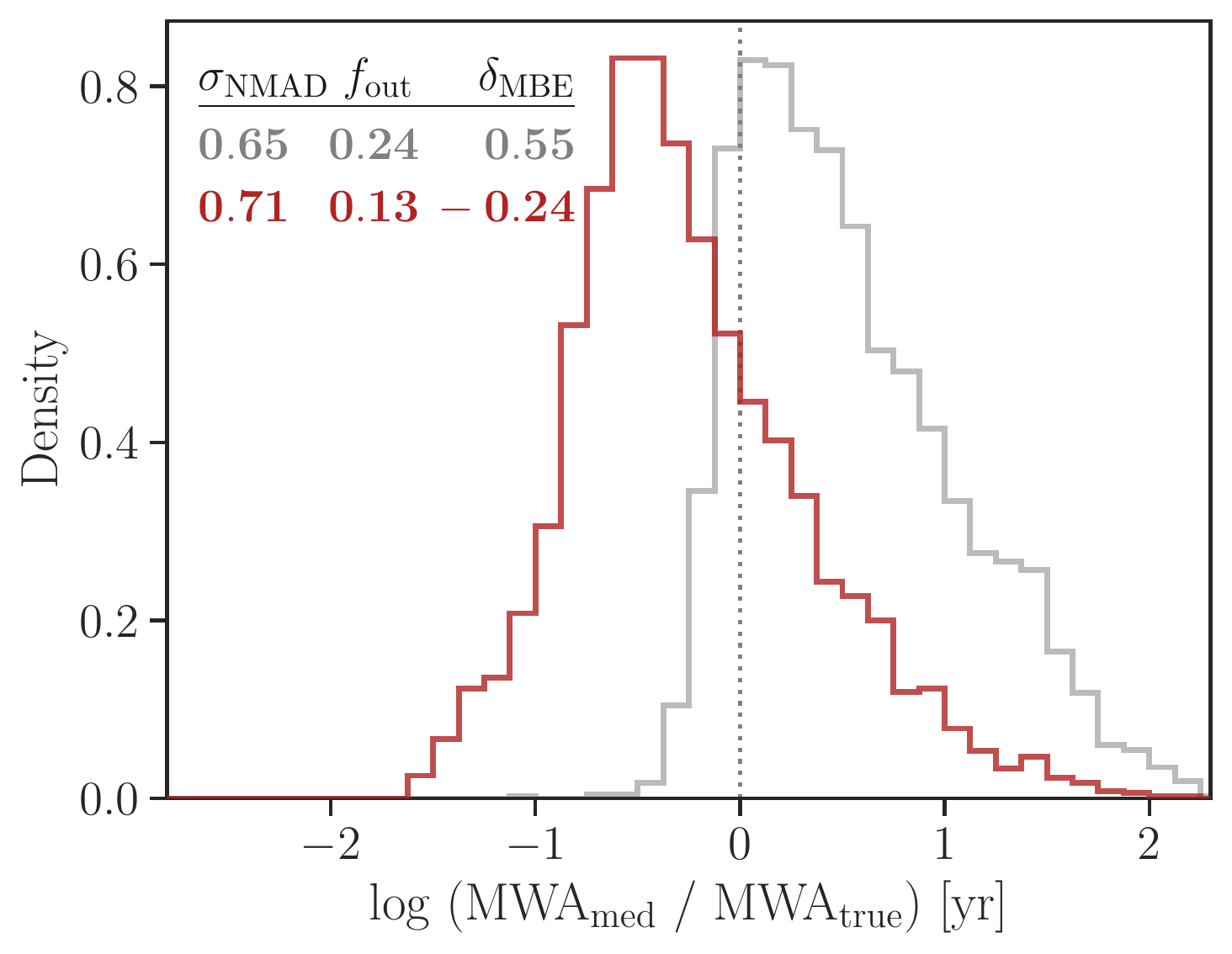}{0.325\textwidth}{}
    \fig{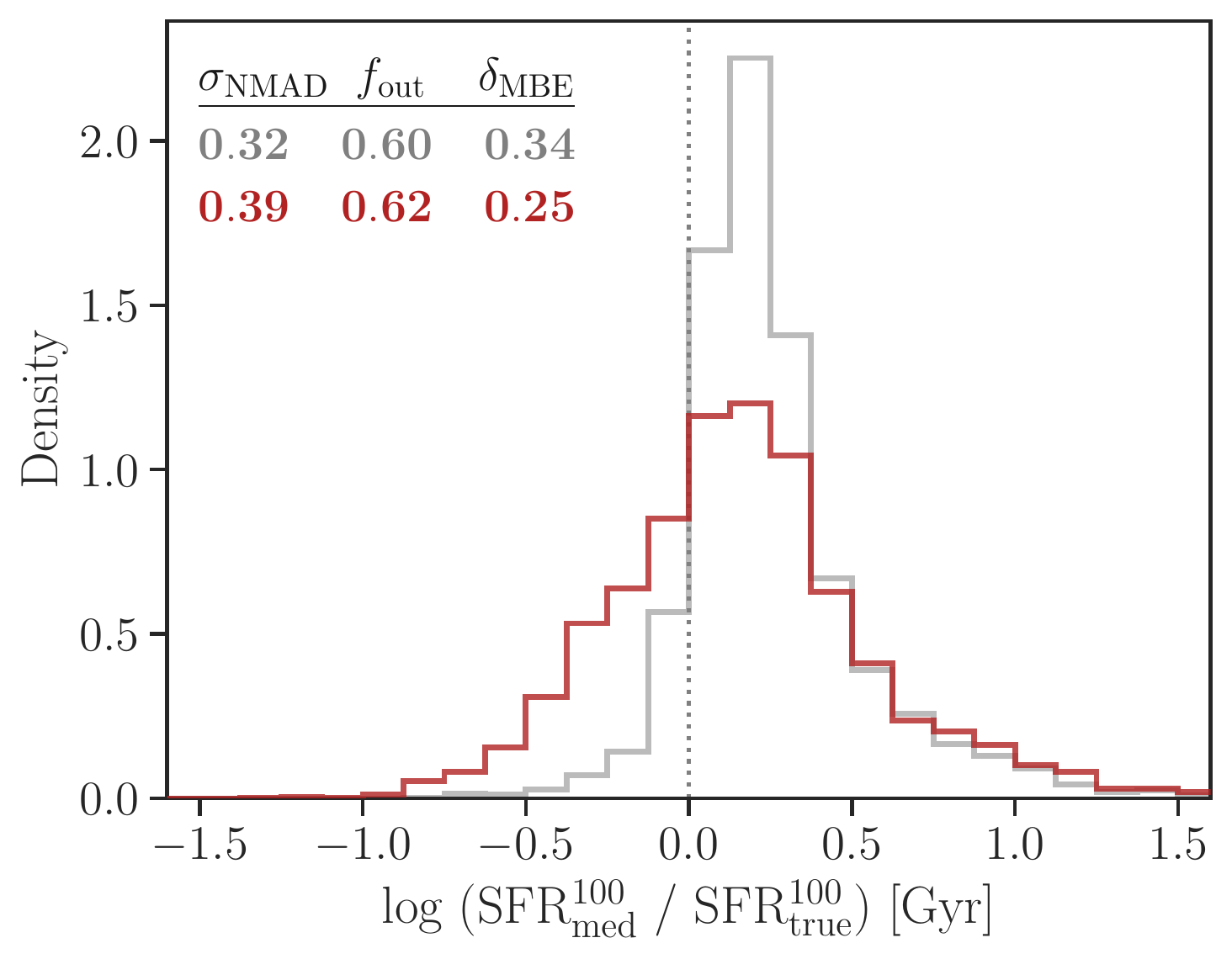}{0.325\textwidth}{}
}
\caption{Here we demonstrate the performance of \pb\ in recovering physical parameters, and contrast it with the fiducial \prospector-$\alpha$ model that assumes uniform priors. Residuals in mass, mass-weighted age, and SFR(0-100Myr) are shown as histograms. Results based on uniform priors are shown in gray, while those based on \pb\ are shown in red. NMAD, outlier fraction, and MBE are reported in the same colors as the corresponding histogram. To avoid counting the error in redshift twice, we only include residuals for the subset where $ (z_{\rm med} - z_{\rm true}) / (1+z_{\rm true}) \leq 0.1$. We find that \pb\ results in less bias consistently in mass, age, and SFR as compared to uniform prior cases.\label{fig:hist}}
\end{figure*}

In this section, we report the parameter recoveries using the \pb\ priors, and contrast them with the uniform priors. We also compare to the redshift inferred via the vanilla \eazy\ templates to put the redshift recovery in context of previous results in the literature. The medians of posterior distributions are quoted in all analyses.

We show histograms of residuals for redshift in Figure~\ref{fig:hist_zred}, and key physical parameters in Figure~\ref{fig:hist}. The scatter in residuals is quantified using the normalized median absolute deviation (NMAD) given its advantage of being less sensitive to outliers than root-mean-square. It is defined as
\begin{equation}
	\sigma_{\rm NMAD} = 1.48 \times {\rm median |\Delta \theta|},
\end{equation}
where $\Delta \theta =  \theta_{\rm med} - \theta_{\rm true}$, with one exception: wherever we evaluate the accuracy of redshift inference, we replace $\Delta \theta$ with $\Delta z = (z_{\rm med} - z_{\rm true}) / (1+z_{\rm true})$. We additionally quantify an outlier fraction, $f_{\rm out}$, in which we define a catastrophic outlier as one with $\Delta \theta > 0.1$, and bias calculated using the mean bias error (MBE) as
\begin{equation}
	\delta_{\rm MBE} = \frac{1}{n} \sum_{i=1}^{n} \Delta \theta.
\end{equation}

\subsection{Assessing parameter recovery \label{subsec:acc}}

We start by evaluating the accuracy of redshift inference, then proceed to key stellar populations metrics. Results on redshifts inferred from the mock catalog assuming uniform priors, and the proposed priors of this letter are shown in the upper panels of Figure~\ref{fig:hist_zred}. Redshifts estimated from \eazy\ are over-plotted as a reference point. The widths of the central distributions in parameter inference of \pb\ and \eazy\ exhibit no significant differences, as evident from their NMAD values. It is reassuring to see that \pb\ and \eazy\ share similar performance, although one discrepancy may be noticed. There exists a small cluster around $\Delta z/(1+z) \sim 0.7$ in the \eazy\ redshifts. This is probably a manifestation of the survey geometry of the mock catalog, which is mostly comprised of young galaxies at $z>1$ with a rising ultraviolet/optical slope due to nebular emission and young stars. \eazy\ tends to mistake these for a rising near-infrared slope of old/quiescent galaxies at $z=0.1-0.3$. \prospector\ recovers the correct redshift by correctly putting star-forming galaxies at high $z$, since its fiducial model already installs a hard limit that no stars can be older than the age of the universe at each time step.
On the other hand, \eazy\ just fits linear combinations of templates. Its default template set is optimized for wider/shallower surveys in which more old quiescent galaxies are present. Adapting a new set of $z$-dependent templates in \eazy\ is likely to alleviate this problem (e.g., the recent set {\verb|sfhz|} which will be described in Brammer et al., in prep.). \pb\ performs understandably better than \eazy\ in this special survey geometry because it is informed by priors based on galaxy evolution that low-mass, high-$z$ galaxies are more common and highly star-forming.

Similar trends are observed in the lower panels of Figure~\ref{fig:hist_zred}, where we test our model on data from a JWST DD program. This perhaps serves as an even more convincing piece of evidence that our approach can perform as least as well as \eazy\ in the context of JWST surveys,
with the added bonus of joint constraints on key stellar populations metrics.

Figure~\ref{fig:hist} demonstrates that the priors proposed in this letter result in a more accurate recovery of mass, mass-weighted age (MWA), and SFR. This is straightforward to conclude from the smaller absolute MBEs because our model is informed by our expectation from galaxy evolution. Reducing the scattering requires more informative data; it is thus natural that we see no significant changes in NMADs.
The recovery of the rest of the parameters in Table~\ref{tab:theta} shows no material difference between difference cases, which is reasonable given that we place no additional prior on them. Discussions on how well these parameters can be constrained within the \prospector\ framework can be found in, e.g.,
\citet{Lower2020,Lower2022,Tacchella2022a}.

Losing the HST coverage spanning $0.4 - 1\mu$m in observed frame does not deteriorate our model performance, beside a 0.1 dex increase in MBE of the redshift recovery. This is because the HST data tend to be associated with larger uncertainties, which means that the likelihood is mostly affected by the JWST bands.

\subsection{Age-mass-redshift degeneracy\label{subsec:deg}}

\begin{figure*}
\gridline{
    \fig{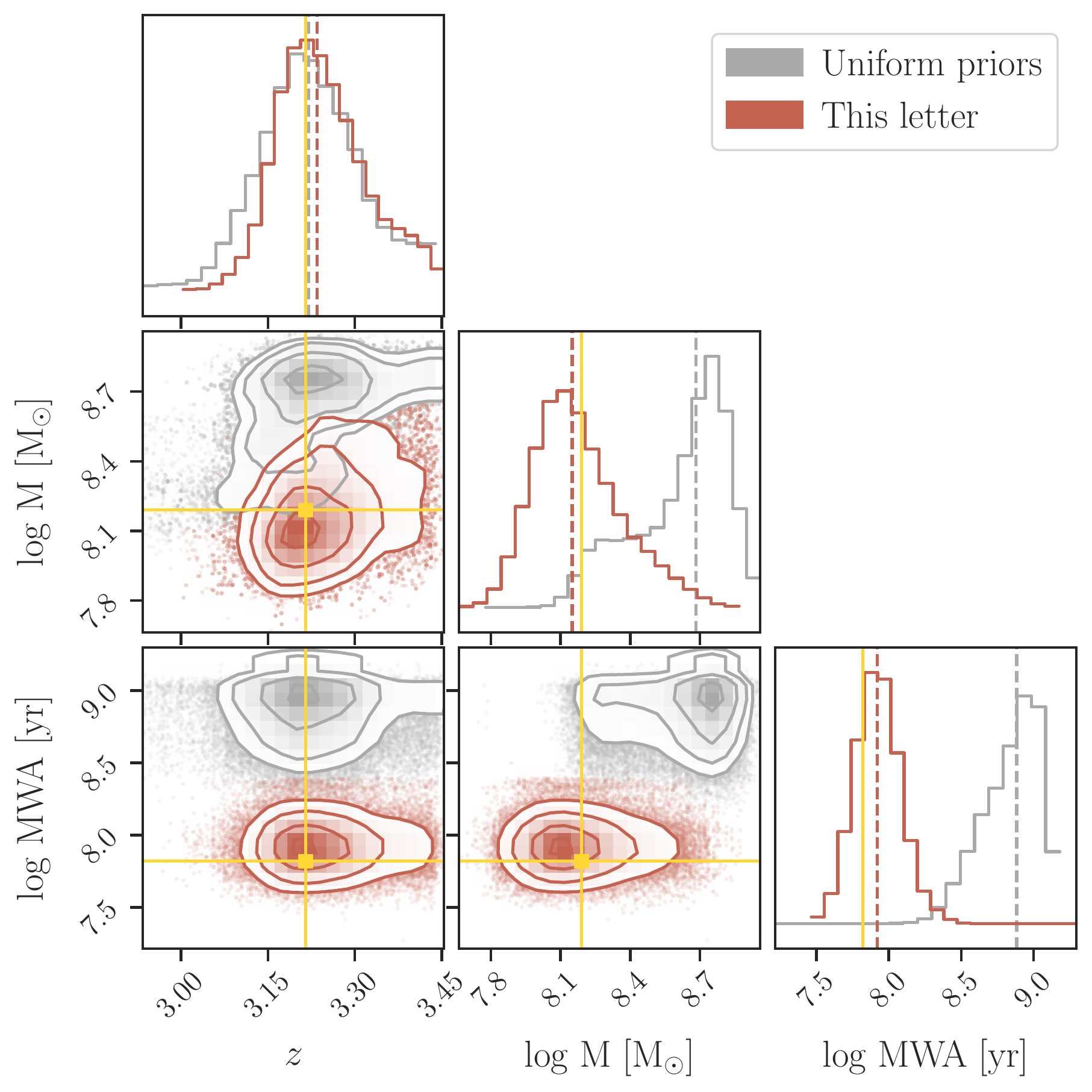}{0.48\textwidth}{(a)}
    \fig{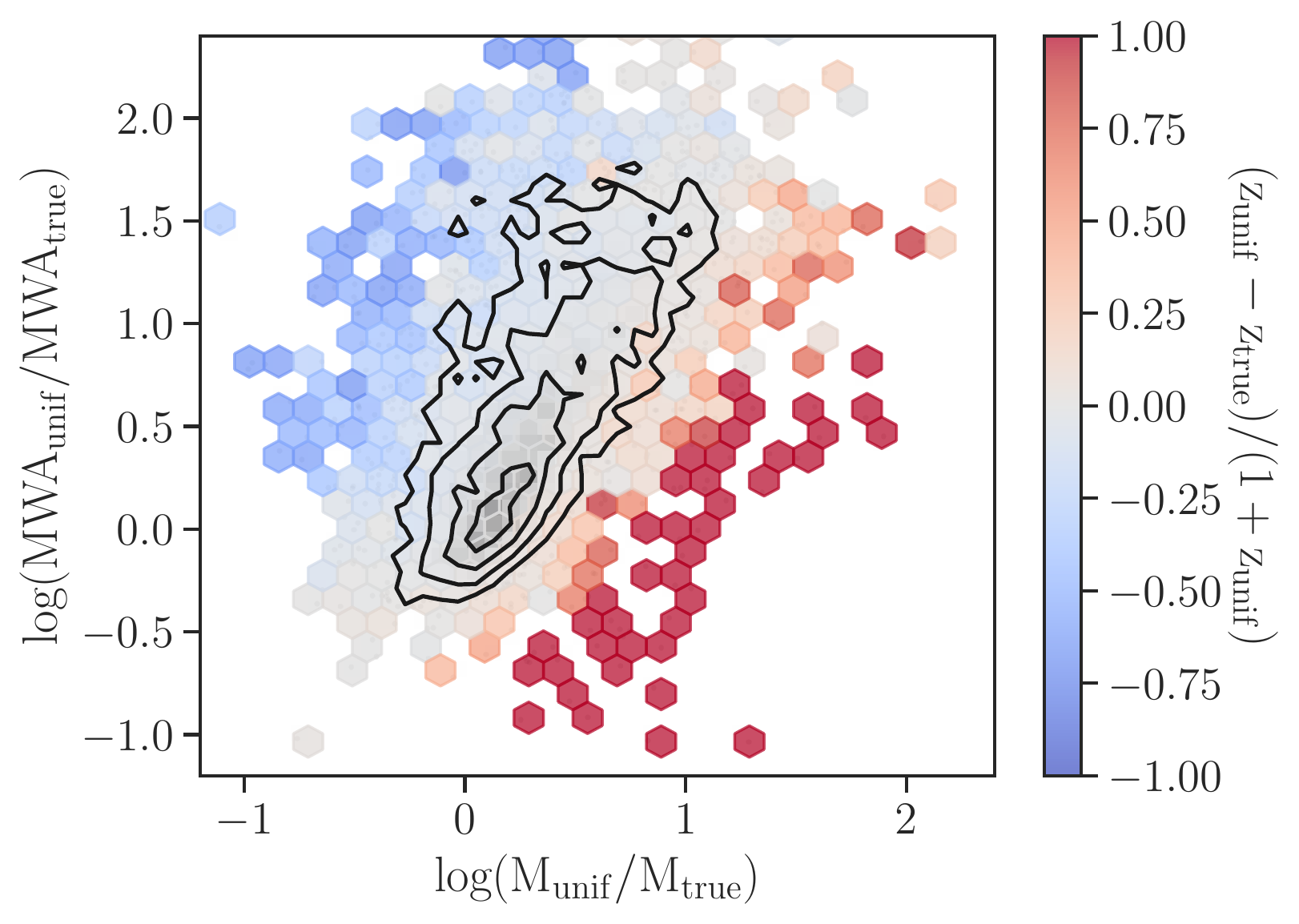}{0.48\textwidth}{(b)}
}
\caption{(a) Posteriors of a sample fit illustrating the degeneracy. The red contours show results assuming the \pb\ priors, while the gray contours show those assuming uniform priors. Truths are over-plotted on the corner plot in yellow. (b) The 2D histogram shows the average redshift offset when assuming uniform priors. Over-plotted are contours showing the density of points. The fact that the contour lines are right along the line of the age-mass-redshift degeneracy suggests that uniform priors get the correct redshifts by exploiting this degeneracy.\label{fig:mcmc}}
\end{figure*}

What is surprising is that the different choices of \prospector-$\alpha$ and \pb\ priors lead to comparable redshift recovery despite the aforementioned improvements in recovered age, mass, and SFR.
Upon examining cases where the uniform priors recover true redshifts, one example of which is shown as Figure~\ref{fig:mcmc}(a), we believe that the flat priors work by exploiting an age-mass-redshift degeneracy: an over-estimation in age leads to a galaxy having a higher mass-to-light ratio; then mass must be over-estimated to produce the same luminosity; however redshift stays the same since color is mostly unaffected. This argument is supported by Figure~\ref{fig:mcmc}(b). The degeneracy is so strong that the flat priors can mostly get the redshift right despite of wrong mass or age. We further discuss the significance of this finding in the following section.

\section{Discussion and Conclusions\label{sec:concl}}

We aim to improve the SED fitting performed on less informative data by including well-motivated priors based on observations of galaxy evolution.
This letter presents the construction, behavior, and influence of three joint priors on mass, galaxy number density, and dynamic SFH. The priors are incorporated into the \prospector\ inference framework. As a result, we are able to better recover stellar mass, age, and SFR. Below we discuss the key findings in this work.

First, we identify and characterize the age-mass-redshift degeneracy. This is likely a general feature of galaxy SEDs (e.g., \citealt{Nagaraj2021}), and in this analysis, it manifests as an deceivingly good recovery of redshift using fiducial \prospector\ model that assumes uniform priors.

Second, our \pb\ model breaks the above age-mass-redshift degeneracy. This has important ramifications. Estimating masses and SFHs at high-$z$ is crucial to many studies such as establishing the onset of galaxy formation and the reionization history. Breaking this degeneracy means that we now have a framework in which redshift, mass, SFH, and other contributions coming from stars, AGNs, gas, and dust can be accurately and self-consistently derived.
An immediate gain is that we can use joint distributions on all parameters from \prospector\ to fit populations models that include full redshift uncertainties. For instance, it becomes possible to improve the mass function methodology from \cite{Leja2020} via the propagation of non-Gaussian redshift uncertainties.

Thirdly, we compare \prospector\ to \eazy\ redshifts. \cite{Leistedt2022} also infers photo-$z$ using \prospector-$\alpha$ and compares to \eazy; in this work, they find that a Bayesian hierarchical model is required to achieve similar performance in estimating redshift as \eazy. The fact that the fiducial \prospector\ model already shows comparable results in Figure~\ref{fig:hist}(a) unambiguously demonstrates the influence of the age-mass-redshift degeneracy. Furthermore, the advantage of encoding galaxy evolution into the inference process is also made visible by the more accurate redshifts from \pb. The less satisfying performance of \eazy\ is likely due to the fact that we deploy the vanilla \eazy\ templates on simulations from a deep but small field, in which massive galaxies are rare. The default \eazy\ setting may become insufficient since it is optimized for a more traditional set-up where old quiescent galaxies constitute a significant proportion. Including redshift dependence into the templates will likely alleviate this problem (Brammer et al., in prep.). We do not carry out a comprehensive comparison to \eazy\ in further detail as the intention here is merely to have a well-tested code to act as a point of reference. We additionally emphasize that the better redshift recovery of \pb\ demonstrates another advantage of our approach: modifying one's model based on a particular survey renders unnecessary once galaxy evolution is encoded in the inference process.

Fourthly, we would like to highlight the simplicity of our SFH prior, which is motivated by downsizing in galaxy formation, and creates rising histories early in the universe and falling histories late in the universe.
Even though the known behavior of SFH($M,z$) is fairly complex, we deliberately choose a simple parameterization. There is no particular reason to tune this in detail to the JAGUAR mock universe, which itself may not be a faithful representation of the true universe. Empirical constraints on SFH at high $z$ remain weak, so we think that such simple approach is preferred, allowing data to inform the inference process to the greatest extent. It is feasible to develop a full population model for $0 < z < 1$ galaxies (e.g., \citealt{Alsing2022}), and constrain this hyper-parameter with data. We leave this improvement for future studies.

Finally, a word of caution: our tests are performed on a mock catalog built for a deep JWST survey with a small area, which means that our mock does not test massive galaxies well since those are rare in this survey. We hope to calibrate the proposed model soon on real data, and also in broader surveys.

To conclude, high-$z$ surveys in the era of JWST increasingly require us to infer more physical information from less informative data. Applying \pb\ to early JWST observations will downweight solutions that are disfavored by current observational constraints on galaxy formation, and hence increase our confidence in the results of our fitting (e.g., \citealt{Labbe2022,Nelson2022}).

This work is thus a timely addition to address the many challenges being put forth by JWST surveys. In the near future, we plan to apply \pb\ to UNCOVER---an ultra-deep Cycle 1 JWST survey targeting first-light galaxies \citep{Bezanson2022}. The data will be an even more stringent test on our model, and will provide a path for future improvements.

\section*{acknowledgments}
We thank the anonymous referee for the constructive comments.
BW is supported by the Institute for Gravitation and the Cosmos through the Eberly College of Science. BW and JL gratefully acknowledge funding from the Pennsylvania State University’s Institute for Computational and Data Sciences through the ICDS Seed Grant Program. Support for Program number JWST-GO-02561.022-A was provided through a grant from the Space Telescope Science Institute under NASA contract NAS5-03127.
Computations for this research were performed on the Pennsylvania State University's Institute for Computational and Data Sciences' Roar supercomputer.
This publication made use of the NASA Astrophysical Data System for bibliographic information.

\facilities{HST(ACS,WFC3), JWST(NIRCam, NIRSpec)}
\software{Astropy \citep{2013A&A...558A..33A,2018AJ....156..123A}, Corner \citep{2016JOSS....1...24F}, Dynesty \citep{Speagle2020}, EAzY \citep{Brammer2008}, Matplotlib \citep{2007CSE.....9...90H}, NumPy \citep{2020Natur.585..357H}, Pandeia, Prospector \citep{Johnson2021},  SciPy \citep{2020NatMe..17..261V}}


\section*{appendix}

\begin{figure}
\gridline{
    \fig{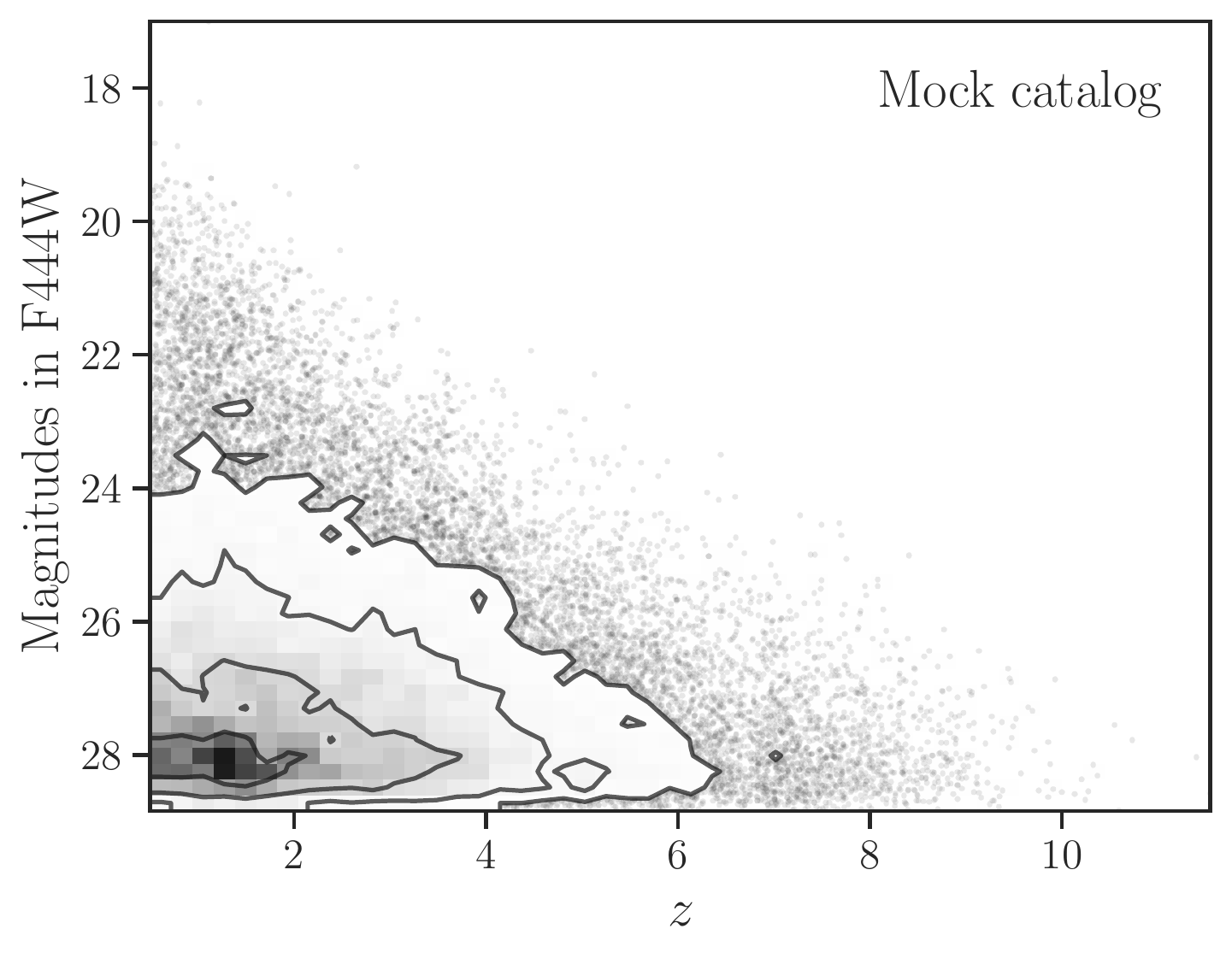}{0.46\textwidth}{}
}
\caption{Magnitudes in F444W are plotted as a function of redshifts for the the mock galaxies that we test our model on. \label{fig:f444w}}
\end{figure}

We supply further details on the properties of our mock catalog in this appendix. The simulated photometry consists of 7 HST and 7 JWST bands spanning $\sim 0.4 - 5 \mu$m in observed frame listed in Table~\ref{tab:depth}, in which we also summarize the assumed depths in each filter. We also show the magnitude in F444W as a function of redshift in Figure~\ref{fig:f444w}.

The noise for the mock observations in the JWST bands are generated using the JWST ETC as follows. First, the uncertainty within a 0.16\arcsec\ radius aperture is calculated given the planned total exposure time in each filter. A flat, faint $f_{\nu}$ source and the point-spread function (PSF) enclosed energy fraction curve is used to convert the flux from $\rm e^{-} s^{-1}$ to nJy. This per-filter aperture uncertainty is then converted to a total flux uncertainty for each object using a per-object filter aperture correction that is determined using 2D Sersic profiles convolved with the PSF. This total flux uncertainty for each band is then used directly as the mock filter flux uncertainty, $\delta f_{\mathrm{X}}$. Noise is randomly drawn assuming a normal distribution with a standard deviation equal to $\delta f_{\mathrm{X}}$, and added to the intrinsic fluxes to yield the mock flux $f_{\mathrm{X}}$.

As for the noise in the HST bands, we similarly assume Gaussian distributions. The standard deviations are estimated from available HST observations overlapping the JWST DD program.

\begin{deluxetable}{ccl}
\tablecaption{Depth per Filter in the Mock Catalog\label{tab:depth}}
\tablehead{
\colhead{Filter} & \colhead{5$\sigma$ depth [AB mag]} & \colhead{Instrument}
}
\startdata
F435W & 26.49 & HST ACS (WFC)\\
F606W & 26.62 & HST ACS (WFC)\\
F814W & 27.16 & HST ACS (WFC)\\
F105W & 25.51 & HST WFC3 (IR)\\
F115W & 26.48 & JWST NIRCam\\
F125W & 26.16 & HST WFC3 (IR)\\
F140W & 25.77 & HST WFC3 (IR)\\
F150W & 28.35 & JWST NIRCam\\
F160W & 29.02 & HST WFC3 (IR)\\
F200W & 28.79 & JWST NIRCam\\
F277W & 28.62 & JWST NIRCam\\
F356W & 28.62 & JWST NIRCam\\
F410M & 28.23 & JWST NIRCam\\
F444W & 28.65 & JWST NIRCam\\
\enddata
\end{deluxetable}

\bibliography{nz_wang.bib}

\end{document}